\PassOptionsToPackage{dvipsnames,table}{xcolor}
\documentclass[sigconf,screen]{acmart}

\usepackage{amsmath,amsfonts}
\usepackage{algorithmic}

\usepackage{csquotes}
\usepackage{graphicx}

\usepackage[dvipsnames,table]{xcolor}

\usepackage{multicol,booktabs}
\usepackage{longtable}

\usepackage{hyperref} 
\usepackage{cleveref}
\usepackage{tikz}
\usepackage{float}
\usepackage{listings}
\usepackage{xspace}
\usepackage{wrapfig}
\usepackage[T1]{fontenc}
\usepackage{courier}
\usepackage{array}

\usepackage{float}
\usepackage{tabularray}

\usepackage[super]{nth}

\usepackage{color}
\definecolor{deepblue}{rgb}{0,0,0.5}
\definecolor{deepred}{rgb}{0.6,0,0}
\definecolor{deepgreen}{rgb}{0,0.5,0}
\newcolumntype{Y}{>{\centering\arraybackslash}X}

\usepackage{listings}
\usepackage{syntax}

\usepackage{expl3,xparse}
\ExplSyntaxOn
\NewDocumentCommand \lstcolorlines { O{green} m }
{
 \clist_if_in:nVT { #2 } { \the\value{lstnumber} }{ \color{#1} }
}
\ExplSyntaxOff

\makeatletter
\let\old@lstKV@SwitchCases\lstKV@SwitchCases
\def\lstKV@SwitchCases#1#2#3{}
\makeatother
\usepackage{lstlinebgrd}
\makeatletter
\let\lstKV@SwitchCases\old@lstKV@SwitchCases

\lst@Key{numbers}{none}{%
    \def\lst@PlaceNumber{\lst@linebgrd}%
    \lstKV@SwitchCases{#1}%
    {none:\\%
     left:\def\lst@PlaceNumber{\llap{\normalfont
                \lst@numberstyle{\thelstnumber}\kern\lst@numbersep}\lst@linebgrd}\\%
     right:\def\lst@PlaceNumber{\rlap{\normalfont
                \kern\linewidth \kern\lst@numbersep
                \lst@numberstyle{\thelstnumber}}\lst@linebgrd}%
    }{\PackageError{Listings}{Numbers #1 unknown}\@ehc}}
\makeatother

\usepackage[most]{tcolorbox}

\usepackage{multirow}
\usepackage{tabularx} 

\definecolor{ghostwhite}{rgb}{0.95, 0.95, 0.95}
\definecolor{really-light-gray}{rgb}{0.96, 0.96, 0.96}

\crefname{lstlisting}{listing}{listings}
\Crefname{lstlisting}{Listing}{Listings}
\usepackage{siunitx}
\newcommand{\nquestions}{\num{67189} }
\newcommand{\naccepted}{\num{27547} }
\newcommand{\nAfterFilter}{\num{13740} }

\newcommand{\nrelevant}{\num{228} }

\newcommand{\collectedpapers}{\num{18729} }
\newcommand{\npapersrelevants}{\num{18} }

\newcommand{\ncategories}{\num{12} }
\newcommand{\ncategoriesaddressed}{\add{\num{23}} }
\newcommand{\ncategoriesnotaddressed}{\add{\num{27}}\xspace}
\newcommand{\nsubcategories}{\num{50} }

\usepackage{pifont}

\definecolor{redwhite}{rgb}{0.93, 0.85, 0.85}
\definecolor{purplewhite}{rgb}{0.85, 0.85, 0.93}
\definecolor{yellowwhite}{rgb}{0.97, 0.97, 0.70}
\definecolor{orangewhite}{rgb}{0.97, 0.7, 0.35}
\definecolor{tropicalrainforest}{rgb}{0.0, 0.46, 0.37}

\newcommand{\keyword}[1]{\tikz[baseline=(char.base)]{
        \node[fill=ghostwhite,rounded corners=.1cm,] (char) {
            {\small#1}
        };
    }\xspace
}

\newcommand{\circled}[1]{\tikz[baseline=(char.base)]{\node[shape=circle,fill,inner sep=2pt] (char) {\textcolor{white}{\textbf{#1}}};}}

\usepackage{xparse}
\usepackage{ifthen}
\usepackage{cprotect}
\usepackage{etoolbox}

\newcounter{cA}
\ExplSyntaxOn
\NewDocumentCommand{\refquestions}{m}{
    $^{(
    \clist_set:Nn \l_tmpa_clist {#1}
    \int_step_inline:nn {\clist_count:N \l_tmpa_clist} {
        \parserefquestions{\clist_item:Nn \l_tmpa_clist {##1}}
        \int_compare:nNnT {##1} < {\clist_count:N \l_tmpa_clist} {,\,}
    }
    )}$
}
\ExplSyntaxOff

\NewDocumentCommand{\parserefquestions}{m}{%
{\underline{
\href{https://answers.ros.org/question/#1}{\thecA}\stepcounter{cA}}}
}

\tcbuselibrary{breakable}
\tcbset{
  width=\linewidth,
  halign=justify,
  center,
  breakable,
  colback=white,
}

\newcommand{\thickhline}{%
    \noalign {\ifnum 0=`}\fi \hrule height 1pt
    \futurelet \reserved@a \@xhline
}

\usepackage{soul}

\usepackage{adjustbox}

\definecolor{PaleGreen}{RGB}{206, 237, 194}
\definecolor{PaleRed}{RGB}{250, 207, 217}
\definecolor{GreenPlus}{RGB}{32, 88, 13}
\definecolor{RedMinus}{RGB}{156, 30, 59}

\newcommand{\showchanges}{0} 

\DeclareRobustCommand{\add}[1]{%
    \ifnum\showchanges=1%
        {{\color{GreenPlus}\mbox{[+]}} \color{GreenPlus} #1}
    \else
        {#1}%
    \fi
}

\DeclareRobustCommand{\rem}[1]{\ifnum\showchanges=1{{\color{RedMinus}\mbox{[-]}} \color{RedMinus} #1}\else{}\xspace\fi
}

\definecolor{dkgreen}{rgb}{0,0.5,0}
\definecolor{dkred}{rgb}{0.5,0,0}
\definecolor{gray}{rgb}{0.5,0.5,0.5}

\lstdefinestyle{javastyle} {
  language=html,
  showspaces=false,
  showtabs=false,
  tabsize=2,
  breaklines=true,
  showstringspaces=false,
  breakatwhitespace=true,
  commentstyle=\color{dkred},
  stringstyle=\color{dkgreen},
  keywordstyle=\color{blue},
  ndkeywordstyle=\color{red},
  basicstyle=\footnotesize\ttfamily,
  numberstyle=\ttfamily\footnotesize\color{gray},
  numbers=left,
  stepnumber=1,    
  firstnumber=1,
  numberfirstline=true,
  numbersep=10pt,
  escapechar=@, 
  xleftmargin=.23in,
  morekeywords={command, node, param},
}

\AtBeginDocument{%
  }

\setcopyright{rightsretained}
\acmDOI{10.1145/3650212.3680350} 
\acmYear{2024}
\copyrightyear{2024}
\acmISBN{979-8-4007-0612-7/24/09}
\acmConference[ISSTA '24]{Proceedings of the 33rd ACM SIGSOFT International Symposium on Software Testing and Analysis}{September 16--20, 2024}{Vienna, Austria}
\acmBooktitle{Proceedings of the 33rd ACM SIGSOFT International Symposium on Software Testing and Analysis (ISSTA '24), September 16--20, 2024, Vienna, Austria}
\acmSubmissionID{issta24main-p719-p}
\received{2024-04-12}
\received[accepted]{2024-07-03}

\begin{document}

\title{Understanding Misconfigurations in ROS: An Empirical Study and Current Approaches}

\author{Paulo Canelas}
\email{pasantos@andrew.cmu.edu}
\orcid{0000-0002-0154-8989}
\authornote{Also with LASIGE, University of Lisbon, Portugal.}
\affiliation{%
  \institution{School of Computer Science\\Carnegie Mellon University}
  \streetaddress{5000 Forbes Ave}
  \city{Pittsburgh}
  \state{PA}
  \country{USA}
  \postcode{15213}
}
\author{Bradley Schmerl}
\orcid{0000-0001-7828-622X}
\email{schmerl@cmu.edu}
\affiliation{%
  \institution{School of Computer Science\\Carnegie Mellon University}
  \streetaddress{5000 Forbes Ave}
  \city{Pittsburgh}
  \state{PA}
  \country{USA}
  \postcode{15213}
}
\author{Alcides Fonseca}
\orcid{0000-0002-0879-4015}
\email{amfonseca@fc.ul.pt}
\affiliation{%
  \institution{LASIGE, University of Lisbon}
  \streetaddress{Campo Grande 016}
  \city{Lisboa}
  \country{Portugal}
  \postcode{1749-016}
}
\author{Christopher S. Timperley}
\orcid{0000-0002-9785-324X}
\email{ctimperley@cmu.edu}
\affiliation{%
  \institution{School of Computer Science\\Carnegie Mellon University}
  \streetaddress{5000 Forbes Ave}
  \city{Pittsburgh}
  \state{PA}
  \country{USA}
  \postcode{15213}
}

\renewcommand{\shortauthors}{Canelas et al.}

\begin{abstract}
The Robot Operating System (ROS) is a popular framework and ecosystem that allows developers to build robot software systems from reusable, off-the-shelf components.
Systems are often built by customizing and connecting components via configuration files.
While reusable components theoretically allow rapid prototyping, ensuring proper configuration and connection is challenging, as evidenced by numerous questions on developer forums.
Developers must abide to the often unchecked and unstated assumptions of individual components.
Failure to do so can result in misconfigurations that are only discovered during field deployment, at which point errors may lead to unpredictable and dangerous behavior.
Despite misconfigurations having been studied in the broader context of software engineering, robotics software (and ROS in particular) poses domain-specific challenges with potentially disastrous consequences. 
To understand and improve the reliability of ROS projects, it is critical to identify the types of misconfigurations faced by developers. 
To that end, we perform a study of ROS Answers, a Q\&A platform, to identify and categorize misconfigurations that occur during ROS development. 
We then conduct a literature review to assess the coverage of these misconfigurations by existing detection techniques.
In total, we find \ncategories high-level categories and \nsubcategories sub-categories of misconfigurations. Of these categories, \ncategoriesnotaddressed are not covered by existing techniques. 
To conclude, we discuss how to tackle those misconfigurations in future work.
\end{abstract}

\ccsdesc[500]{Computer systems organization~Robotics}
\ccsdesc[300]{General and reference~Empirical studies}

\keywords{ROS, Misconfigurations, Empirical Study, Literature Review}

\maketitle


\section{Introduction}
\label{sec:introduction}

The Robot Operating System (ROS), known as the \enquote{Linux of Robotics}, is the de facto open-source framework for building robot software~\cite{quigley2009ros}.
ROS's package ecosystem provides developers with reusable, off-the-shelf components that implement common robot functions (e.g., perception, planning, localization, drivers)~\cite{estefo2019ros,SophiaICSME20}.
In theory, ROS allows developers to quickly prototype robot software by integrating such components and adjusting their parameters via configuration files (e.g., Launch XML, ROS Param YAML) to match their intended application and environment.

Despite the relative ease of integrating components using configuration files in ROS, \textit{correctly} configuring systems presents a considerable challenge.
For instance, the software's configuration depends on the robot's hardware and operating environment (e.g., specific types of sensors and their placement in the robot).\footnote{\url{https://answers.ros.org/question/59087}}
When communicating with each other, components must make matching assumptions about their environment (e.g., when the camera is 10 cm above the wheels, a reference transformation is required).\footnote{\url{https://answers.ros.org/question/227092}}
Finally, timeliness properties must be considered when integrating components to avoid negatively impacting robot behavior (e.g., frames may be dropped when processing high-resolution image streams, leading to unstable and dangerous motion).\footnote{\url{https://answers.ros.org/question/248656}}

To provide plug-and-play functionality, ROS components are typically required to make assumptions about the context in which they are used (e.g., a topic should receive messages of the correct type at a certain frequency) that are neither checked nor documented~\cite{estefo2019ros}.
Misconfigurations occur when one or more components make different, conflicting assumptions about the robot, leading to unintended and potentially dangerous behavior (e.g., property damage, human harm) during deployment.
Given the importance of safety within this domain, it is vital to identify misconfigurations before the robot is deployed, and as early as possible.
To that end, the robotics software engineering community has begun to develop tools to detect certain misconfigurations, such as those related to 
physical units~\cite{kate2018phys}, architecture~\cite{timperley2022rosdiscover}, and reference frames~\cite{kate2021physframe}.

To systematically tackle the misconfiguration problem, it is critical to understand the types of misconfigurations that occur in the wild and whether existing tools are designed to detect them.
Based on our own experiences with ROS, we know that physical units, architectural, and reference frames are not ROS's only categories of misconfiguration.
Software misconfigurations have been thoroughly studied in different contexts of software development (e.g., security~\cite{rahman2023security,eshete2011security}, databases~\cite{liao2018configuration,yin2011misconfig}, cloud computing~\cite{uchiumi2012misconfig,zhang2014encore,huang2015confvalley} and networks~\cite{mahajan2002networkmisconfig}).
However, misconfigurations within ROS are inherently different due to their cyber-physical nature.

In this work, we set out to \textbf{identify the broader set of misconfigurations that impact ROS systems and to determine which detection techniques address them and which misconfiguration types are going undetected}.
This knowledge can guide future research in the robotics software engineering community in developing novel tools and techniques to address them. 

We first derive a taxonomy of misconfigurations within ROS systems by conducting an empirical study of relevant questions posted to ROS Answers, a ROS-specific Q\&A site similar to Stack Overflow.
Secondly, we determine the extent to which state-of-the-art analysis tools help to address those misconfigurations by
conducting a literature review of analysis papers published at several major software engineering, architecture, testing, and robotics conferences.
Finally, as part of our analysis, we highlight misconfigurations unaddressed by existing techniques and further discuss research opportunities in developing new techniques for them.

Through our study, we make the following contributions:

\begin{itemize}
    \item A taxonomy of misconfigurations in ROS, based on a qualitative study of a popular Q\&A platform (\Cref{sec:taxonomy});
    \item A literature review of the state-of-the-art approaches and how they cover the misconfigurations (\Cref{sec:lit-review});
    \item A dataset of misconfigurations and questions manually analyzed and categorized that can be used to guide future studies and develop novel techniques (\url{https://zenodo.org/doi/10.5281/zenodo.12642380}).
\end{itemize}

\section{\add{Background}}
\label{sec:background}

In this section, we provide a high-level introduction to ROS as a middleware for building component-based robotics software and as an open-source ecosystem of reusable components.

Systems in ROS are built as a collection of independent processes, known as \emph{nodes} or \emph{components}, each responsible for providing certain functions (e.g., perception, planning, control, driver interfacing).
At its core, ROS's responsibility is to provide the \enquote{plumbing} that facilitates communication between distributed components.
The bulk of communication within ROS follows an anonymous publish/subscribe pattern~\cite{santos2017mining}.
At one end, components (e.g., a camera driver) publish messages to named topics (e.g., \lstinline{/camera/color/image_raw}).
On the other hand, components (e.g., object detection) subscribe to those same topics to receive messages.
Neither the publishers nor the subscribers are directly aware of one another's identities (i.e., they are spatially decoupled~\cite{eugster2003pubsub}).
Moreover, communications are defined at run-time via calls to the ROS API (e.g., strings are used to state topics by name).

One of ROS's biggest strengths is its rich ecosystem of generic open-source components that can be reused for common robot functions~\cite{SophiaICSME20,macenski2023survey}.
For example, MoveIt!~\cite{coleman2014moveit} provides motion planning and execution for manipulation,
\texttt{ros\_control}~\cite{roscontrol} provides implementations of low-level controllers (e.g., velocity, joints, effort),
and \texttt{ros_localization}~\cite{robotlocalization} provides filters for state estimation (e.g., by fusing GPS and IMU data).\
By reusing off-the-shelf components, developers can, in theory, reduce the cost and complexity of building robot software. 
However, as these components are generic, they must be configured to work in specific contexts.

ROS uses configuration files to customize and arrange those components into a functioning ensemble.
These include Launch XML (or Python) files, which launch and compose each component within the system, and ROS Parameter YAML files, which are used by components at run-time to customize their behavior (e.g., specifying a color format and topic name for camera images).
General-purpose components use these configuration files to tailor their behavior to a particular robot, application, or environment.
Particularly complex and variable components and subsystems (e.g., Nav2, MoveIt!, ros\_control) go beyond providing a fixed set of parameters and embed a limited domain-specific language within the ROS parameter system.

Together, these aspects of ROS enable rapid prototyping, encourage component reuse (e.g., exchange components without modifying the rest of the system), and reduce the cost and complexity of building robots.
However, due to ROS's dynamic, spatially decoupled architecture, developers must follow assumptions and conventions when integrating components into their system, which are neither enforced nor documented~\cite{afzal2020challenges, estefo2019ros}.
Failing to do so leads to \emph{misconfigurations}. 
Often, misconfigurations do not produce meaningful errors and are uncovered via manual and laborious debugging of erroneous behaviors at run time.
\section{Study of Misconfigurations}
\label{sec:taxonomy}

\begin{figure*}[ht]
    \includegraphics[width=\linewidth]{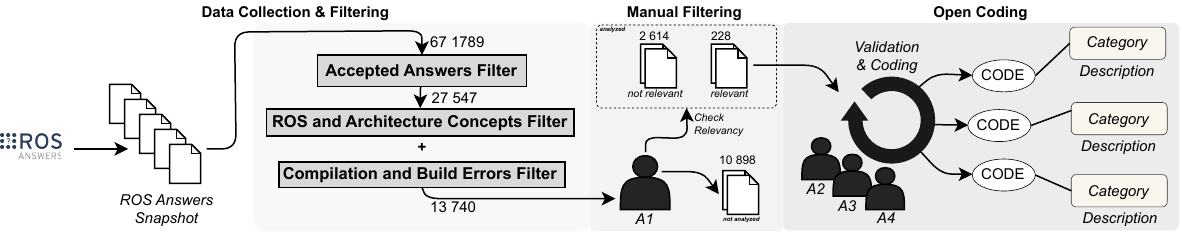}
    \centering
    \caption{
        Three-step methodology for analyzing ROS Answers questions.
        \textbf{Step 1} collects a snapshot of \nquestions questions.
        \textbf{Step 2} selects questions with accepted answers containing ROS or software architecture concepts and filters questions about compilation or building issues.
        \textbf{Step 3} produces a sample of \nrelevant questions filtered based on their relevance. 
        Questions are divided into stages, where their codings are iteratively improved.
    }
    \label{fig:eval-methodology}
\end{figure*}

To understand which misconfigurations current tools can address, we first need to know the different types that exist.
Therefore, we ask the following research question:

\textbf{RQ1: What kinds of misconfigurations do developers make when building robot software systems with ROS?}

To answer this question, we perform an empirical study of ROS Answers, the, until recently, primary Q\&A forum for ROS~\cite{estefo2019ros}.\footnote{In August 2023, Open Robotics decided to move to the Robotics Stack Exchange (\href{https://robotics.stackexchange.com}{robotics.stackexchange.com}).}
Q\&A platforms are designed for users to post their problems, including explaining the scenario where the problem occurs. 
While only some of the questions pertain to misconfiguration, we found many such instances in a prospective search.
Q\&A websites provide more detailed misconfiguration examples than social coding platforms (e.g., GitHub) since commit messages often lack important context, making it difficult or impossible to reliably identify misconfigurations.
Furthermore, issue trackers often describe bugs in individual components rather than the difficulties of integrating those components into a working system.
To that end, we perform a thematic analysis of questions posted to ROS Answers.

In the rest of this section, we describe our methodology (\Cref{sec:misconfigurations:methodology}) and its associated threats to validity (\Cref{sec:misconfigurations:threats}) before presenting our taxonomy of ROS misconfigurations (\Cref{subsec:results-taxonomy}).

\subsection{Methodology}
\label{sec:misconfigurations:methodology}

\Cref{fig:eval-methodology} outlines our high-level methodology, which takes inspiration from studies of similar Q\&A platforms~\cite{albergo2022xacro,tian2019archsmells}.
Below, we describe each step of our methodology.

\begin{figure}[t]
\centering
\begin{tikzpicture}
    \node[anchor=south west,inner sep=0] at (0,0) {\includegraphics[width=\linewidth]{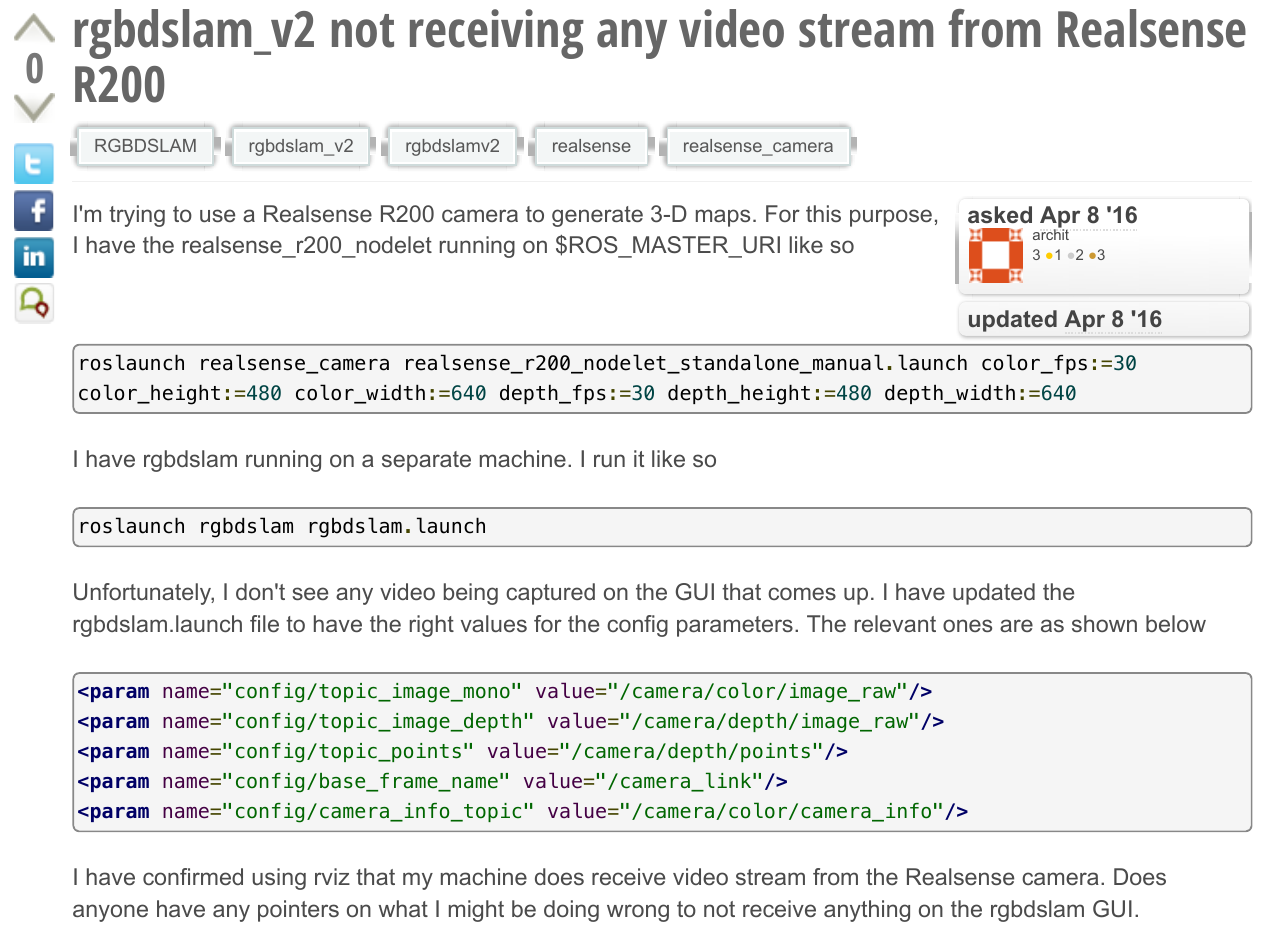}};
    \draw[black,thick,fill=black] (8.1,5.5) circle (0.2cm);
    \node at (8.1,5.5) {\color{white}\small{\textbf{1}}};
    \draw[black,thick,fill=black] (8.1,1.2) circle (0.2cm);
    \node at (8.1,1.2) {\color{white}\small{\textbf{2}}};
    \node[anchor=south west,inner sep=0] at (0,-2.1) {\includegraphics[width=\linewidth]{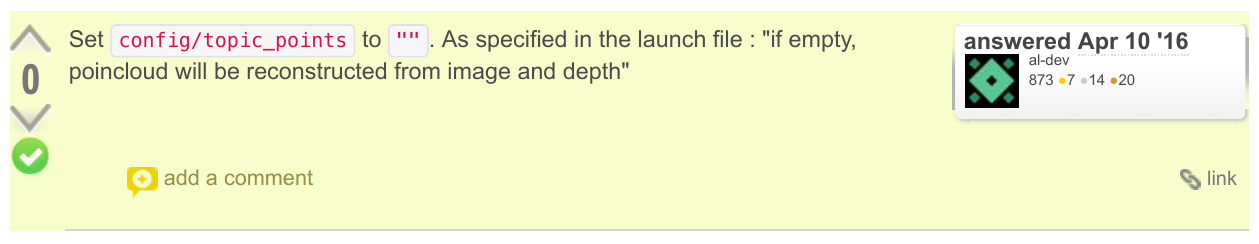}};
    \draw[black,thick,fill=black] (8.1,-1.65) circle (0.2cm);
    \node at (8.1,-1.65) {\color{white}\small{\textbf{3}}};
\end{tikzpicture}
\caption[Example of a ROS Answers question.]{
Example of a ROS Answers question corresponding to a misconfiguration where the developer incorrectly defined a parameter \add{value}.
Each question contains a title ({\scriptsize\circled{1}}), content with text and source code ({\scriptsize\circled{2}}), and metadata about the author, date, and number of votes.
Questions may include comments, answers, and an accepted answer ({\scriptsize\circled{3}}).
}
\label{fig:ros-answers-question}
\end{figure}
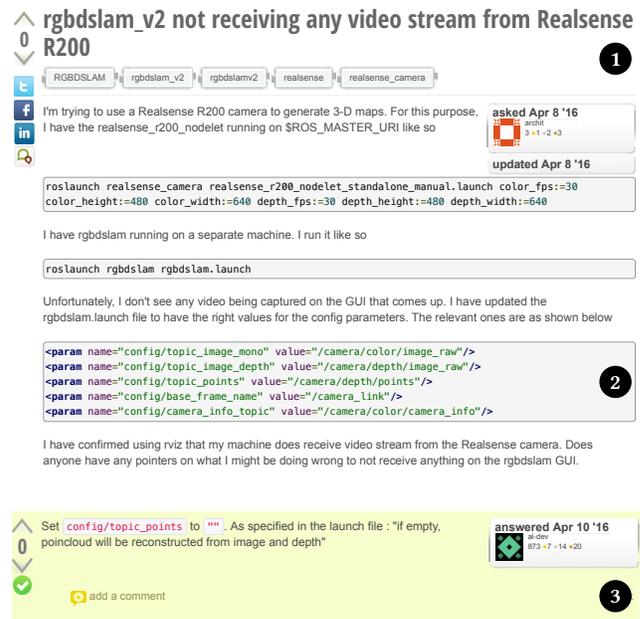

\textbf{Data Collection \& Filtering. }%
\label{sec:misconfigurations:data}%
We first gathered all \nquestions questions posted to ROS Answers between January \nth{1}, 2011, and November \nth{20}, 2022.
\Cref{fig:ros-answers-question} presents an example of a question (\underline{\href{https://answers.ros.org/question/231458}{231458}}) and its accepted answer.
We then filtered these questions to a set of \naccepted by selecting only those with an accepted answer as they offer an alternative perspective and accepted solution from the user.

During the second step, we narrowed the accepted questions to those referring to ROS and architectural concepts, expecting to obtain misconfiguration questions.
Reducing the set of questions is a common practice in the literature~\cite{albergo2022xacro,tian2019archsmells,albonico2021mining}.
We used common ROS concepts as defined in the \href{http://wiki.ros.org/ROS/Concepts}{ROS Wiki}:
node, subscribe topic, message, parameter, service, action, launch, publish, and subscribe.
Since developers indirectly define their system's architecture  when changing configurations, we also identify questions related to software architecture errors~\cite{garlan2009mismatch}:
architecture, mismatch, assumption, incompatibility, inconsistency, integration, and configuration.
Subsequently, the set was further refined by removing questions containing keywords related to installation and build errors (e.g., \texttt{build error} or \texttt{compil*}).
\add{Build errors are comparatively easy to determine and diagnose, as developers can examine the error messages generated. 
This work focuses on undetected misconfigurations during the deployment process, subsequently impacting robots behaviors. 
These errors are significantly more challenging to detect and trace back to their source.}
This final filtering step provided a total of \nAfterFilter accepted questions considered for sampling.

\begin{figure*}[t]
\centering
\includegraphics[width=\linewidth]{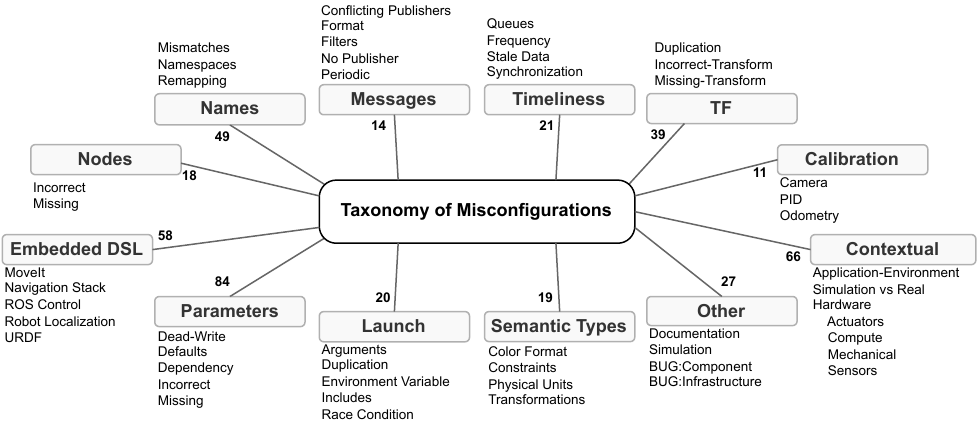}
\caption{Mindmap of the misconfigurations identified from the study presenting the \ncategories high-level categories of misconfigurations, and their \nsubcategories sub-categories level. Each misconfiguration contains the number of questions annotated with the code. Each question may refer to more than one misconfiguration.
}
\label{fig:taxonomy}
\end{figure*}

\textbf{Manual Filtering.}
The first author (A$_1$) randomly selected questions from the \nAfterFilter questions and labeled them as relevant or irrelevant by analyzing their content to determine if it described a failed attempt to configure the system \add{(i.e., interacting with the configurations or source code files)}.

\textbf{Open Coding.}
The first author (A$_1$) iteratively provided sets of relevant questions to three \add{other} authors (A$_2$, A$_3$, A$_4$) who applied an open coding~\cite{ezzy2013qualitative} approach to construct a taxonomy of misconfigurations.
At each step, the authors individually proposed an updated set of codes for their sub-set of questions before discussing those codes and merging them into a revised taxonomy.
The subsets were constructed such that each question was analyzed by three different authors, allowing a diversity of perspectives to be captured and reducing the author bias.
This process continued until reaching the saturation point~\cite{glaser2017discovery} (i.e., no further changes were made to the taxonomy after reaching the end of a step) after analyzing \nrelevant questions, resulting in a final taxonomy of \ncategories categories and \nsubcategories subcategories of ROS misconfiguration.

\textbf{Labeling.}
Finally, we labeled each of the \nrelevant questions using the final taxonomy.
Half of the questions were labeled by one pair of authors (A$_1$, A$_2$) and the other half by a different pair (A$_3$, A$_4$).
We calculated the agreement by dividing the number of codes both authors agreed on by the total number of codes used.
This first step led to an agreement of 84.12\% and 85.5\% for each pair of authors.
Then, each pair compared the codes that differed by one code and adjusted their code if in agreement.
Furthermore, to determine documentation-related questions, the authors collected all questions annotated with documentation, discussed random instances of these, and re-annotated all questions until they reached an agreement on using this code.
Finally, the authors discussed 26 questions with initial disagreements.
The authors who did not analyze a given question were arbiters during the discussion.

\subsection{Threats to Validity}
\label{sec:misconfigurations:threats}

\textbf{External Validity.}
We identify two primary external validity threats: the generality of our results to (a) different ROS versions and distributions and (b) expert users. 
The first threat relates to the possible predominance of ROS 1 over ROS 2 questions and the impact of ROS distributions.
Given the relatively recent release of ROS 2~\cite{foote2022ros2} in 2018, we expected more questions related to ROS 1.
We found that the analyzed questions rarely specified ROS or distribution versions (37 out of 228), making it infeasible to determine version-specific misconfigurations. 
\add{Furthermore, as there are few architectural differences between ROS versions, we believe that our findings generalize to both versions.}
\rem{Nevertheless, most of our taxonomy generalizes to both versions, so our findings apply to both ROS versions and distros.}
The second threat concerns the applicability of our findings to real-world scenarios.
By sampling data from a popular ROS Q\&A platform, we are addressing real-world developer issues.
Nevertheless, we recognize that industrial settings may present unique, undisclosed misconfigurations.
Our taxonomy provides a basis for further studies in such contexts.

\textbf{Internal Validity.}
We identify four main threats to internal validity: 
the initial samping method, the generalizability of the sampled data, biases in question analysis, and potential misrepresentation of misconfiguration types in the sampled ROS Answers questions.
The first threat regards the initial sampling and validation step with only one author, possibly introducing personal biases in the selection.
To mitigate this step, we performed a preliminary study similar to the current one, in which all authors looked at a sample of questions and validated them as relevant or not relevant.
To address the second threat, we sampled relevant ROS Answers questions, inspected, validated, and categorized them until reaching theoretical saturation, preventing the introduction of unaddressed categories with new questions.
For the third threat, we iteratively analyzed questions and validated their relevance and misconfiguration categories with at least two authors.
The forth threat concerns the sample's representativeness.
We focus on questions related to ROS concepts and software architecture, which are more likely to contain misconfigurations.
While other misconfigurations may be overlooked, this does not invalidate the identified categories.

\subsection{Results}
\label{subsec:results-taxonomy}

In this section, we describe each high-level category and sub-category of \keyword{Misconfiguration} in detail, along with relevant examples. 
\Cref{fig:taxonomy} depicts the mindmap of the taxonomy of misconfigurations.

\textbf{Messages.}
Most ROS communication occurs via messages exchanged between components over named topics. \rem{(i.e., publish-subscribe).}
Through our analysis, we identified five misconfigurations related to messaging.

\keyword{Format} misconfigurations arise due to a mismatch between two or more components in the message format that are exchanged on a shared topic.
For instance, the \textit{initialpose} topic, representing the initial position and orientation of the robot in the map, accepts \textit{geometry\_msgs/Pose} messages. Both publishers and subscribers must respect the message format when exchanging messages on this topic.
However, a misconfiguration occurs when either end breaks the \enquote{contract} and expects a different type of message.

\begin{figure}[t]
\centering
\includegraphics[width=\linewidth]{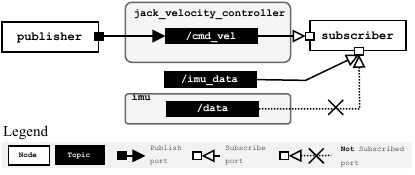}
\caption{
    Example of a \textit{Name Mismatch} from ROS Answers, where the developer mistyped the subscriber's topic name.
    The subscriber expects data \add{from \texttt{imu/data}}, but due to the wrong connection \add{to \texttt{imu\_data},} no data is received.
}
\label{fig:error-names-mismatch}
\end{figure}

Components may expect a specific number of publishers to the topics to which they subscribe.
This assumption eases components' expectations of the frequency of messages they receive.
\keyword{No Publisher}
misconfigurations occur when a component subscribes to a topic no publisher sends messages to.
The subscriber waits indefinitely for messages that never arrive.
\Cref{fig:error-names-mismatch} presents an example where the subscriber never receives messages due to a topic name mismatch.
\keyword{Conflicting Publishers}
misconfigurations appear when there is more than one simultaneous publisher to a topic that should only be accessed by a single publisher.
This leads to messages that may provide opposite, conflicting instructions.

The message\_filters API is used to filter incoming messages on a given topic (i.e., messages satisfying a given condition trigger a callback) and synchronize messages across multiple topics (e.g., invoke a callback once data is received from multiple sensors).
In particular, we noticed difficulties in using this API (\keyword{Filters}) to synchronize two topics without losing messages.
Finally, \keyword{Periodic} misconfigurations occur when the correct system execution relies on messages being periodically published at a specific frequency (e.g., camera, lidar, IMU data).
Misconfigurations occur when a component stops publishing data continuously and other parts of the system continue to wait, indefinitely, to receive that data.
For instance, a pedestrian detection node must consistently publish images regarding pedestrian's position estimates to function correctly.


\textbf{Launch.} 
Launch files are the primary means of orchestration within ROS, used to launch and glue together individual components with specific configurations into the system.

Developers often use \add{the ability to import other launch files recursively} to improve modularity and simplify reuse by writing individual launch files for separate sub-systems (i.e., a collection of components that work together to perform tasks such as perception, planning, or control).
When writing a launch file for an entire system, developers introduce a layer of abstraction, requiring them to only reason about what launch files to include rather than worrying about configuring every individual node and subsystem.
We observed cases where the developer either \keyword{Includes} inappropriate launch files for the given context or fails to include crucial launch files necessary for the robotic system's functioning.

We also observe node \keyword{Duplication} errors in launch files, where two or more nodes are instantiated with the same name (e.g., by accidentally launching the same node with the same name).
In this case, ROS complains at run-time that the name is taken, and the second node crashes upon launch.

Since ROS architectures are defined at run-time, launch files are susceptible to \keyword{Race Conditions}.
For instance, nodes may publish messages to topics before the subscriber finishes launching, causing those messages to be lost.
Nodes can also be sensitive to the ordering of \texttt{<node>} and \texttt{<param>} tags within launch files:
In ROS 1, a \texttt{<node>} may launch before the parameters are stored on the global parameter server, leading to parameter misconfigurations.

To allow components to be customized to a particular system, launch files support \keyword{Arguments}, whose values may be provided by the command line, a parent launch file, or a specified default value.
Those can be accessed via string interpolation (e.g., via \texttt{\$(arg name-of-arg)}) within the launch file or through the command line, and are typically used to specify ROS parameter values, rename nodes, control the inclusion of particular nodes, or set the system time.
Arguments are prone to many of the same issues as parameters (e.g., dead writes and unintentional use of default values).

\keyword{Environment variables} are also used to customize the behavior of individual components in launch files.
Misconfigurations can occur when necessary environmental variables for nodes are not specified or when incorrect values are assigned to those variables.

\textbf{Parameters.}
\label{subsec:taxonomy-parameters}
In ROS, parameters are used to adjust the behavior of components to their intended deployment.
Parameter values are typically provided by launch files, which are used to compose multiple components into a functioning ensemble.
\Cref{lst:error-parameter} presents an example of a launch file with two parameter-related misconfigurations where the developer forgets to include a \texttt{tf\_prefix}. \add{This parameter is critical when working with multi-robot systems,} to create separate transform trees for each robot. 
\add{Since the parameter is not defined, the system, by omission, uses the default value, and a single transform tree is used for all the robot systems.}

Since parameters are defined and used at run-time within ROS,
components may unexpectedly crash when a required parameter is \keyword{Missing}, or behave in an unintended manner when the component falls back on a \keyword{Default} value for a missing parameter.
Both of these types of parameter misconfiguration can also be caused by \keyword{Dead Writes} where the wrong name is used to specify a parameter (e.g., due to a typographical error or a name change refactoring).
These cases can be hard to debug as there is no static checking and warnings may not be produced for missing or unused (i.e., dead-write) values.

Misconfigurations can occur when using \keyword{Incorrect} parameter values.
Those values may be universally incorrect (e.g., out of bounds) or contextually incorrect for the given robot, environment, and application.
When defining parameter values, developers also need to be cautious of potential \keyword{Dependency} issues, where the behavior of a given parameter is changed by the value of another parameter (e.g., a parameter that enables or disables a feature).

\begin{figure}[t]
\begin{lstlisting}[language=html,
    caption={Example where \texttt{tf\_prefix} parameter required for operating with multiple robots is \textit{Missing}, leading the system to use the \textit{Default Parameter} value.}, 
    captionpos=b, 
    showspaces=false, 
    escapechar=@,
    label={lst:error-parameter},
    linebackgroundcolor={\lstcolorlines[yellow]{9,10,11}}]
<!-- robot urdf model -->
<param name="robot_description" command="cat $(find urdf_pkg)/urdf/my_robot.urdf" />

<!-- robot state publisher node -->
<node pkg="robot_state_publisher"
    type="state_publisher"
    name="robot_state_publisher">

    <param name="~tf_prefix"
            value="robot_name"
            type="str" />
</node>
\end{lstlisting}
\end{figure}

\textbf{Semantic Types.}
Even when components correctly make assumptions about the message format shared on a given topic, they can still make incorrect assumptions about the message content (i.e., their semantic types).
For instance, two components may correctly exchange a \texttt{sensor\_msgs/Image}, but the publisher sends color images while the subscriber expects grayscale images.

Components implicitly assume that messages satisfy specific \keyword{Constraints} over their contents.
For example, values are within certain bounds (e.g., positions, velocities, motor values), specific coordinate frames are used, or the range of laser scan measurements is respected.
Type-related misconfigurations may stem from the improper use of images and point clouds, such as mismatched assumptions between components about the \keyword{Color Format} of the image or point cloud (e.g., grayscale vs. color images) or an incorrect assumption that all of the images and point clouds that are shared on a given topic have been subject to specific \keyword{Pointcloud Transformations}
(e.g., resizing, compression, or color conversion) or \keyword{Image Transformations}.
Finally, misconfigurations can occur due to mismatched assumptions on the \keyword{Physical Units} of data that are exchanged between components.
As robots interact with the real world, ensuring that the component's physical units match is critical.
For instance, a publisher describes rotational velocity using radians per second, but the subscriber expects the same quantity in degrees per second.


\textbf{Names.}
Every node, topic, service, action server, and parameter within ROS has an associated name specified at run-time either as a field or property within a Launch XML file or as a string in the source code of a component.\footnote{http://wiki.ros.org/Names}
Since names are resolved at run-time, it is easy to introduce \keyword{Mismatches} between two or more components in the name of a resource (e.g., topic, parameter, service, or embedded DSL configurations).
\Cref{fig:error-names-mismatch} illustrates an example of one of these errors from our dataset, where the developer incorrectly defined the topic's name in the subscriber.
\add{Since the name is incorrectly configured, the subscriber receives no messages as there are no publishers for that topic.
Detecting this misconfiguration is challenging, as the topic names only differ by a specific character and are usually only detected during execution when the system does not behave as intended.}
Another instance of a name mismatch relates to typos when writing the configurations for the robot localization. 
For instance, the developer incorrectly defines the name of the \texttt{world\_map}, a parameter specifying the frame treated as a fixed reference frame.

ROS implements a hierarchical naming structure to promote encapsulation. By convention, this system groups related resources and allows multiple instances of the same node to be used simultaneously (e.g., one node for each camera).
ROS's hierarchical naming structure is implemented via namespaces:
Every resource belongs to a namespace, denoted by a forward slash in the name of that resource (e.g.,\texttt{/right_camera/raw_image} belongs to \texttt{/right_camera}). 
Furthermore, namespaces may be stacked (e.g., \texttt{/vehicle_a/right_camera/raw_image}) where \texttt{/} denotes the root of the hierarchy, known as the global namespace.
We observe that \keyword{Namespaces} misconfigurations occur most commonly when developers forget to use namespaces or otherwise use the wrong namespace, leading to naming collisions and causing the system to have an unintended architecture.

In addition to namespaces, ROS also relies on name \keyword{Remapping} to help encapsulation and reuse:
ROS launch files allow the mapping of the name of a certain resource (e.g., topic, service) onto a different name within the context of a particular component.
This ability is used to wire a general-purpose component into the necessary configuration for a specific system.
For instance, when using the \texttt{image_proc/resize} component\footnote{http://wiki.ros.org/image_proc} to resize camera images, users must remap the subscribed \texttt{image} topic (i.e., incoming camera images) and published \texttt{$\sim$image} topic (i.e., resized image) onto appropriate source and destination topics (e.g., \texttt{/right_camera/image_color} and the topic target named \texttt{/right_camera/image_color_resized}).
Incorrect or missing remappings can lead to message loss and unintended behavior.

\textbf{Nodes.}
Nodes (i.e., components) are the processes within ROS that collectively form a working robot by performing computation, exchanging information, and interacting with the robot hardware.
In our study, we observe cases where either crucial components are \keyword{Missing}
or an \keyword{Incorrect} component is used for the particular robot, environment, or intended application.
For example, the developer incorrectly uses a planner, which leads the robot to navigate erroneously, possibly against walls.
A special case for missing a component is a missing nodelet manager.
In this case, developers do not define the qualified manager name, allowing all nodelets to be assigned to that manager.

\begin{figure}[t]
\centering
\includegraphics[width=\linewidth]{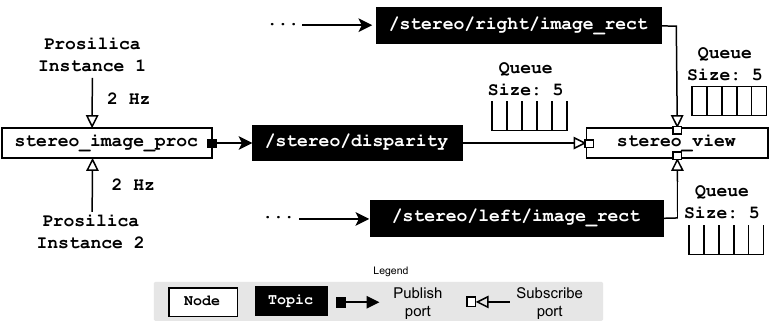}
\caption[Example of a ROS Answers question behavioral misconfiguration.]{
    Example of a misconfiguration \add{where the} developer gets images from \lstinline|stereo_image_proc| at 2Hz, converting them to disparity images.
    The \lstinline|stereo_view| needs matching left, right, and disparity images, but the slow processing speed of disparity images causes left and right images to arrive faster and fill the queue, dropping messages.
    The \lstinline|stereo_view| skips these until a matching of images is available.
}
\label{fig:error-queues-freq}
\end{figure}

\textbf{Timeliness.}
Robots are real-time systems: messages must be sent between components (e.g., control signals, state estimates) by a certain deadline to prevent unintended and dangerous behavior.
Timeliness misconfigurations occur when the timing assumptions of interacting components are mismatched.

Both publishers and subscribers within ROS 1 have associated \keyword{Queues}, which are used to buffer either outgoing or incoming messages.
Determining an appropriate size for this queue is crucial to ensure both timely and correct behavior:
A queue size that is too small can lead to message loss, whereas an overly long queue can lead to excessive compute resource usage and message delays.
For instance, \Cref{fig:error-queues-freq} presents a queue misconfiguration within a robot with stereo vision.
The \lstinline|stereo_view| node requires three images: left, right, and the disparity image (i.e., the difference between left and right).
In the proposed architecture,\footnote{https://answers.ros.org/question/9108} the disparity image is computed using the other images.
However, computing the disparity image takes time and computational resources.
\add{Due to the desynchronization of the sensors,} when \add{the disparity image is} ready, the original left and right images have already been overwritten in their corresponding queues.
To fix this misconfiguration, the developer can increase the queue size to avoid overwriting or throttle the publishing rate of the cameras to account for time taken to produce the disparity image.

This example is an instance of \keyword{Frequency}
misconfiguration between publishers and subscribers, where subscribers expect to receive messages at a given frequency to operate correctly.

To function safely, certain components rely on an uninterrupted stream of data that accurately describes the state of the robot and its environment.
For example, motion planning in a dynamic environment requires timely and accurate estimates of the position of both the robot and potential obstacles in the scene.
Unintended and unsafe behavior can occur when \keyword{Stale Data}
no longer accurately represent the current state of the robot and its environment, as messages are not published at a high enough frequency.

Finally, \keyword{Synchronization}
of certain messages is essential for correct system operation.
For instance, \add{in example to showing the queue misconfiguration,} \Cref{fig:error-queues-freq} presents \add{an example where the sensors used are sources of multiple misconfigurations.
One source of misconfiguration is the lack of synchronization between camera frequencies.} \rem{this type of misconfiguration where camera frequencies are not synchronized.}
There is a mismatch when the stereo\_view processor receives the information from each camera.
Synchronization of the receiving data is required to ensure that all images are consumed simultaneously, keeping the original images available before consuming the corresponding disparity image.

\textbf{TF.}
Robot systems typically rely on a large number of 3D coordinate frames to reason about the relative position and orientation of the robot, its physical parts, and its environment.
\texttt{tf}\footnote{http://wiki.ros.org/tf} is a core ROS library that uses a special tree structure to allow users to transform between coordinate frames (e.g., to determine the position and orientation of the robot's gripper relative to the robot's base).

We observe three major types of TF-related misconfigurations:
\keyword{Incorrect Transform}, 
either due to a typo, a misunderstanding of the transform tree semantics, or a mistake about the geometry of the robot.
\keyword{Missing Transform}, 
where the developer forgets to provide a transformation between a parent and child frame.
Finally, \keyword{Duplication} 
of transforms, where the developer publishes the same TF transform from multiple conflicting sources. (i.e., there should be a single source of truth).


\textbf{Embedded DSL.}
\label{subsec:taxonomy-dsl}
ROS, its associated toolchain, and some of its most popular (and general purpose) packages rely on their own custom configuration formats.
Some ROS packages embed their configuration formats inside of ROS's parameter system, effectively forming an embedded DSL.
In other cases, a standalone file is used (e.g., URDF).
We observed issues related to the use and configuration of the Navigation Stack, \textit{ros_control}, \textit{robot_localization}, MoveIt!, and URDF.
These misconfigurations contain more specific types of issues (e.g., parameter issues).
This high-level category presents the types of configuration files related to these misconfigurations.

\keyword{Navigation Stack}
provides mobile robots with the ability to use odometry and sensor values to localize their position and navigate within a 2D plane by sending velocity commands to the mobile base.\footnote{https://wiki.ros.org/navigation}
Specifically, we saw issues concerning the correct definition of motion planning parameters.
For instance, developers may select an inappropriate planner or fail to adapt the parameters of that planner to the robot and environment.

\keyword{URDF}
(Universal Robot Description Format) files describe robots in terms of their links, joints, transmissions, sensing capabilities, collision geometry, and physical properties (e.g., inertia, contact coefficients, joint dynamics).\footnote{https://wiki.ros.org/urdf}
URDF files are used for visualization, simulation, and motion planning.
For instance, when developers forget to specify the joints between two links or incorrectly provide the robot description of two systems in the same file.

\keyword{MoveIt} is a platform for building manipulators using ROS that incorporates algorithms for motion planning, manipulation, kinematics, control, and navigation~\cite{coleman2014moveit}.
Successfully configuring it for a robot in a particular environment relies on careful configuration of numerous parameters (e.g., planner density, and padding offsets).

\keyword{ROS Control}
allows developers to integrate and compose multiple off-the-shelf control algorithms into their system.
To behave safely and operate as intended, each controller needs to be adapted to the specifics of each robot, which, when done incorrectly, can result in a misconfiguration.

Finally, \keyword{Robot Localization} provides a node collection for performing state estimation (i.e., Kalman filters) and integrating GPS data: they fuse data from multiple sensor sources (e.g., IMUs, GPSs, odometry) to obtain a robust estimate of the robot's state (i.e., position, rotation, velocity).
We observe issues that stem from missing or incorrect transforms, incorrect units, and sensor mismatches (e.g., attempting sensor fusion without wheel or visual odometry).


\textbf{Calibration.}
\label{subsec:taxonomy-calibration}
Robots rely on a suite of sensors to perceive their environment.
To ensure that the robot's understanding of its environment is and remains accurate, those sensors must be calibrated.
In our study, we observed misconfigurations due to the miscalibration of \keyword{Cameras}, 
\keyword{Odometry}, and 
\keyword{PID}
controller parameters, all of which required a manual change.
In all of these cases, the mistake was either (a) forgetting to calibrate entirely (e.g., camera intrinsics and extrinsics), (b) relying on default parameters that were inappropriate for the robot (e.g., PID defaults), or (c) using incorrect values (e.g., wheel radius).

\textbf{Contextual.}
\label{subsec:taxonomy-contextual} %
\add{Misconfigurations can also occur when the system's configuration is tweaked according to a specific context.} A simple example is developing the robot's software within a simulation context and deploying it to a physical robot or vice versa. 
Some configurations and parameters need to be different when changing contexts, leading to \keyword{Simulation vs. Real} misconfigurations when the behavior does not match (typically by not adjusting the necessary configurations).
Furthermore, we encountered \keyword{Application-Environment} misconfigurations where the component's configuration depends on the type of application of the system and the surrounding environment.
For instance, the configurations of the components inside and outside a warehouse can be different due to weather, lighting conditions, and surface grip.

The robot's physical configuration (i.e., its sensors, actuators, mechanical components) naturally imposes restrictions over the set of plausibly correct software configurations; configuring the robot's software therefore requires an understanding of the robot's physical configuration.
We identified four sub-categories where \keyword{Hardware}
components require careful software configuration:
(1) \keyword{Actuators}, 
where the configuration of the parameters for a component in the system depends upon the exact actuators that are used,
(2) \keyword{Sensors}, 
where the type of sensors and their positioning on the system restricts parameter configuration,
(3) \keyword{Mechanical}, 
where the robot's mechanical hardware imposes restrictions of the space of meaningful and safe parameters,
and (4) \keyword{Compute} 
where the configuration impacts resource usage.
Furthermore, robots may have specific hardware limitations that must be addressed via software configuration.
For instance, \add{one source of misconfiguration in} \Cref{fig:error-queues-freq} \rem{contains a} is resource-related, \rem{misconfiguration} where components take longer to process overly large camera images.

\textbf{Other Challenges.}
During the analysis, we encountered questions not related to misconfigurations.
\add{\keyword{Documentation} questions occur due to missing or outdated documentation in ROS.
This category is not considered a misconfiguration as the developer could not progress with the configuration and reach a misconfiguration due to the lack of documentation.}
We highlight this \add{category} as part of our taxonomy since outdated or lack of documentation may cause developers to introduce misconfigurations.
Developers may not know what components to use or how to use them correctly in their systems.
\rem{Secondly,} We \add{also} encountered instances where the issues were related to the component having a \keyword{BUG:Component} rather than
being a configuration issue, or the \keyword{BUG:Infrastructure} in which the system is running contained an error (e.g., RViz).
Finally, we identified cases where the misuse of \keyword{Simulation} prevents the correct functioning of the system.
Developers can use simulation time to playback previously recorded data (e.g., sensor readings) in a time-synchronized manner (e.g.,  during testing and debugging). 
However, developers unfamiliar with the concept of simulation time may forget to define it, resulting in incorrect system behavior (e.g., different times in the receiving of data).

\begin{figure*}[ht]
    \includegraphics[width=\linewidth]{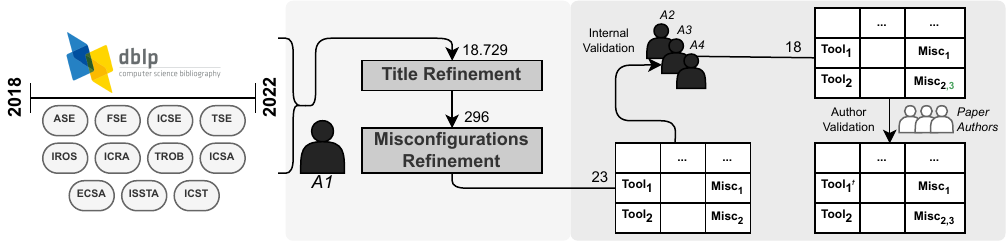}
    \centering
    \caption{
        Methodology of the literature review.
        We collected prior work from 2018 to 2022 from top conferences and journals in software engineering, robotics, architecture, and testing.
        We refined the search by looking at the paper titles, then reading the papers and matching them to misconfigurations.
        We perform an internal validation with three authors and author validation with the authors of each technique to \add{validate} the proposed tool for misconfiguration matching.
    }
    \label{fig:methodology-lit-review}
\end{figure*}
\section{Study of Existing Tools}
\label{sec:lit-review}

After categorizing the different types of misconfigurations in ROS-based systems, it is essential to identify which categories are overlooked when using existing state-of-the-art analysis techniques.
To that end, we address the following question:

\textbf{RQ2: To what extent do current techniques address these categories of misconfiguration?}

Answering this question reveals the gaps in current tooling and helps guide the development of new tools to increase the coverage of misconfigurations that can be detected earlier in the development process, avoiding errors in deployment.
To address this objective, we conducted a literature review following best practices outlined by Snyder et al.~\cite{snyder2019litreview}.
Our search strategy, influenced by Albonico et al.~\cite{albonico2023ros}, focused on sourcing works from the software engineering, software architecture, software testing, and robotics communities, which are the likely places for such techniques to have been reported.
\Cref{fig:methodology-lit-review} illustrates our methodology for performing the literature review, which we describe below.

\subsection{Methodology}
\label{subsec:lit-review-methodology}

Our literature review methodology consists of three stages:

\textbf{Stage 1. Collection:}
Using DBLP,\footnote{\url{https://dblp.org}} we collected all papers of journals and conferences and associated co-located events between 2018--2022 inclusive from the major software engineering (ICSE, FSE, ASE, TSE), software testing (ICST, ISSTA), software architecture (ICSA, ECSA), and robotics (ICRA, IROS, TROB) venues.
We gathered \collectedpapers paper links, titles, venues, and \add{respective} years. \rem{for those papers.}

\textbf{Stage 2. Refinement \& Collection:}
The first author started by manually inspecting paper titles and searching for keywords that describe any of the subcategories of misconfigurations.
Furthermore, for non-robotics venues, we searched in the titles for robotics-related topics.
In contrast, in the robotics venues, we searched for concepts related to verification, testing, and repairing misconfigurations.
Whenever a prior work seemed relevant, we manually inspected it by looking at the abstract, followed by the introduction, approach, and conclusion.

Then, we annotated the relevant papers with the misconfigurations they cover and their type of analysis (static or dynamic).
We consider a technique static if it performs verification without executing the system and dynamic otherwise.
Although general-purpose testing techniques potentially cover all the identified misconfigurations in theory, we only considered those that explicitly cover them in their problem statement or examples provided in their evaluation.
This refinement considered \npapersrelevants relevant papers.

\textbf{Stage 3. Internal \& Authors Validation:}
We performed a two-step validation of the matching each technique and the misconfiguration. We first validated our findings internally with the other authors of this study (A$_2$, A$_3$, A$_4$), experts in software architecture, software engineering, and robotics, to mitigate against missing or incorrect categorizations.
We then asked the authors of each technique to validate our findings externally.
\add{Since ROS Answers questions do not contain executable examples to test each technique's ability to detect a particular kind of misconfiguration, the external author validation helps us validate our assumptions of each tool's ability.}
To perform an external author validation, we gathered the contact details for the authors of each technique and emailed them to asking if our classification is correct and what other misconfigurations, if any, their technique addresses.\footnote{At the time of submission, 10 of the \npapersrelevants techniques were validated by the authors of those techniques.}
We analyzed the answers received by the \add{papers'} authors.
When in disagreement, we compare both mappings and re-analyze the paper.
From the proposed misconfiguration mapping, 3 out of the 10 authors \add{of each technique} proposed an update by adding only 1 extra category.
For all but one of thise addition, we agreed with the correction and updated our categorization.
The sole case where we disagreed with the authors' additional categories was Santos et al.~\cite{santos2021haros}, which allows developers to synthesize run-time monitors and discover bugs by writing properties using the HAROS Property Language (HPL)~\cite{santos2018harospbt}. 
Similar to general-purpose test cases, these properties act as tests that may detect specific instances of misconfiguration but do not cover the overall \emph{category} of misconfiguration.
While this extensibility is essential, as is writing integration tests in general, we consider this and other similar general-purpose testing techniques outside the scope of this literature review as, with the appropriate instrumentation and scaffolding (i.e., test inputs), testing tools can \add{theoretically identify any misconfiguration, even if it is seldom practical.}

\subsection{Threats to Validity}
\label{sec:lit-review:threats}

\textbf{External Validity.}
We identify the generalizability to other venues and the time frame selected as external threats.
This literature review focused on searching for relevant work at top conferences and journals in software architecture, software engineering, software testing, and robotics.
However, our findings may miss relevant tools by focusing on specific conferences and a particular time frame (2018--2022).
Extending the search for further years and conferences and journals presents a challenging task as the intersection of multiple research areas quickly increases the prior work required for manual inspection.

\textbf{Internal Validity.}
We identify the manual inspection of prior work and the subjectivity in categorization as internal threats.
The initial paper refinement was done by only one of the authors, and there is a threat when selecting relevant prior work based on the paper titles where the author may overlook relevant papers.
By searching for terms related to misconfigurations and the verification, testing, and repair of robotic systems, we expect to mitigate this threat.
There is also a threat when performing the mapping of each technique to a misconfiguration.
To mitigate this mapping, each technique was analyzed by two authors, and we tried to validate the mapping with the original paper authors.

\newcommand{\specialcell}[2][c]{%
  \begin{tabular}[#1]{@{}c@{}}#2\end{tabular}}
\begin{table*}[ht]
\rowcolors{2}{gray!15}{white}
\caption{Overview of each technique, the type of analysis (\textbf{D}ynamic or \textbf{S}tatic), and the sub-categories of misconfigurations each addresses. The main categories of misconfigurations are as follows:
\textbf{(Ca)} Calibration, \textbf{(Co)} Contextual, \textbf{(M)} Messages, \textbf{(N)} Names, \textbf{(O)} Other, \textbf{(P)} Parameter, \textbf{(T)} Semantic Types, ${\dagger}$ Author validated, $*$ Authors Disagreement.}
\label{tab:comparison-lit-review}
\begin{tabularx}{\textwidth}{lcccY}
\toprule
\rowcolor{gray!50}

\textbf{Reference} & \textbf{Venue} & \textbf{Year} & \textbf{Analysis} & \textbf{Misconfigurations} \\ \hline 
    Kate et al.~\cite{kate2018phys}$^{\dagger}$                                      & FSE       & 2018 & S & 
    $\vcenter{
        \textbf{(T)} Physical Units}
    $
    \\
    Burgueno et al.~\cite{burgueno2018physical}$^{\dagger}$     & ICSE-RoSE & 2018 & S/D & 
    $\vcenter{
        \textbf{(T)} Physical Units
        \textbf{(T)} Constraints
    }$
    \\
    Witte et al.~\cite{witte2018archros}                     & ICSE-RoSE & 2018 & S/D& 
    $\vcenter{
        \textbf{(N)} Mismatches
        \textbf{(M)} No Publisher
    }$
    \\
    Wuest et al.~\cite{wuest2019online}                   & ICRA & 2019 & D & 
    $\vcenter{
        \textbf{(P)} Missing
    }$
    \\
    Cramariuc et al.~\cite{cramariuc2020learning}                   & ICRA & 2020      & D & 
    $\vcenter{
        \textbf{(Ca)} Camera
        \textbf{(Co)} Sensors
    }$
    \\
        Carvalho et al.~\cite{carvalho2020harosafety}$^{\dagger}$   & IROS & 2020 & D & 
    $\vcenter{
        \textbf{(T)} Constraints
        \textbf{(M)} Format
        \textbf{(M)} No Publisher
    }$                          
    \\
    Wigand et al.~\cite{wigand2020environment}               & IROS & 2020 & S & 
    $\vcenter{
        \textbf{(Co)} Actuators
        \textbf{(Co)} Application-Environment
        \textbf{(Co)} Mechanical
    }$                          
    \\
    Kate et al.~\cite{kate2021physframe}                         & FSE       & 2021 & S  & 
    $\vcenter{
        \textbf{(TF)} Incorrect Transform
        \textbf{(TF)} Missing Transform
    }$
    \\
    Jung et al.~\cite{jung2021swarmbug}$^{\dagger}$                              & FSE       & 2021 & D  & 
    $\vcenter{
        \textbf{(P)} Incorrect
        \textbf{(P)} Dependency
        \textbf{(P)} Defaults
        \textbf{(Co)} Application-Environment
    }$
    \\
    Kortik et al.~\cite{kortik2021llt}                                 & ICRA      & 2021 & S/D    & 
    $\vcenter{\specialcell{
        \textbf{(M)} Format
        \textbf{(M)} No Publisher 
        \textbf{(M)} Conflicting Publishers}
    }$
    \\
    Santos et al.~\cite{santos2021haros}$^{\dagger*}$                        & ICSE-RoSE & 2021 & S/D & 
    $\vcenter{
        \textbf{(N)} Mismatches
        \textbf{(M)} No Publisher
    }$                            
    \\
    DeVries et al.~\cite{devries2021analysis}$^{\dagger}$                                                  & ICSE-SEAMS & 2021 & S/D & 
    $\vcenter{
       \textbf{(Co)} Application-Environment
        \textbf{(Co)} Mechanical
        \textbf{(Co)} Sim-vs-Real
    }$
    \\
    Taylor et al.~\cite{taylor2022sa4u}$^{\dagger}$                            & ASE & 2022 & S & 
    $\vcenter{\specialcell{
        \textbf{(T)} Physical Units
        \textbf{(TF)} Incorrect Transform 
        \textbf{(TF)} Missing Transform}
    }$
    \\
    Kim et al.~\cite{kim2022robofuzz}$^{\dagger}$                      & FSE & 2022 & D  & 
    $\vcenter{\specialcell{
        \textbf{(DSL)} URDF
        \textbf{(P)} Incorrect
        \textbf{(Co)} Sim vs Real \\
        \textbf{(Co)} Sensors
        \textbf{(Co)} Actuators
        \textbf{(BUG)} Infrastructure}
    }$
    \\
    Das et al.~\cite{das2022calibration}                            & ICRA & 2022 & D   & 
    $\vcenter{
        \textbf{(Co)} Sensors
        \textbf{(Ca)} Camera
    }$
    \\
    Heiden et al.~\cite{heiden2022pis}$^{\dagger}$                                & ICRA & 2022 & D  & 
    $\vcenter{
        \textbf{(O)} Simulation
    }$
    \\
    Timperley et al.~\cite{timperley2022rosdiscover}$^{\dagger}$   & ICSA & 2022 & S & 
    $\vcenter{\specialcell{
        \textbf{(M)} Conflicting Publishers
        \textbf{(M)} No Publisher 
        \textbf{(M)} Format \\
        \textbf{(N)} Mismatches
        \textbf{(N)} Remapping 
        \textbf{(P)} Dead Write
        \textbf{(P)} Incorrect
        }
    }$
    \\
    Han et al.~\cite{han2022lgdfuzzer}  & ICSE & 2022 & D &
    $\vcenter{
       \textbf{(P)} Incorrect
       \textbf{(P)} Dependency
    }$
    \\
\bottomrule
\end{tabularx}%
\end{table*}

\subsection{Results}
\label{subsec:lit-review-results}

\Cref{tab:comparison-lit-review} presents the mapping between the techniques and the misconfigurations.
We identify the misconfigurations each technique addresses, the type of analysis it performs, static or dynamic, and the venue.
We identified techniques that statically or dynamically detect misconfigurations, although some tools are static techniques whose verification is optionally extended with dynamic analysis.
For instance, Burgueno et al.~\cite{burgueno2018physical} statically analyzes physical units described by a modeling language and generates model invariants checked during the execution of the system.

We also identified techniques that do not detect misconfigurations but rather automatically infer or optimize configuration parameters.
The inference of the configurations helps prevent misconfigurations, as developers do not configure the system manually.
For example, Wuest et al.~\cite{wuest2019online} automatically infers geometric and inertia parameters, and Heiden et al.~\cite{heiden2022pis} probabilistically infers simulation parameters.

For each technique, we describe the misconfigurations it addresses with an important caveat.
Although a technique addresses a specific misconfiguration, it does not mean it is solved.
Some techniques address particular misconfigurations for specific robotic systems (e.g., Swarmbug) and contain limitations.
For instance, Phys~\cite{kate2018phys} and SA4U~\cite{taylor2022sa4u} both detect physical unit misconfigurations.
\add{While Phys performs static analysis, allowing it to detect physical unit errors before execution, SA4U requires execution information to detect the misconfigurations.}

Overall, we identified \npapersrelevants related works that address \ncategoriesaddressed of \nsubcategories sub-categories of misconfigurations.
Parameters, Messages, and Contextual are the  misconfigurations most addressed by current techniques.
\add{On the other hand, no misconfigurations related to the Timeliness, Nodes, and Launch categories are currently addressed, and within the Embedded DSL category, only the URDF dialect is considered in current techniques.}
\section{Related Work}
\label{sec:related}

In this work, we studied the types of misconfigurations that developers face and what techniques can address them.
As the presented misconfigurations are not specific to any ROS version or distribution,
we expect our findings to generalize to the many ROS systems.

Prior work studied types of bugs in robotic systems and autonomous vehicles.
For instance, \num{27,25}\% of the bugs in autonomous vehicles (AV) software detected are misconfigurations~\cite{garcia2020autonomous}.
Instances of the misconfigurations we encountered were also found in Unmanned Aerial Vehicles (UAV)~\cite{wang2021uavbugs}.
\num{19.6}\% of the bugs encountered in a study of two UAVs, PX4 and Ardupilot, are misconfigurations such as parameter misuse and missing, parameter limits related to hardware, and inconsistencies related to sensors and libraries.
Our taxonomy of misconfigurations not only addresses ROS-specific issues but also other types of misconfigurations related to the cyber-physical nature of these systems.

Similar to the misconfiguration we encountered, the simulation to real-world transition is challenging in the services robotics domain~\cite{garcia2020robotics}.
Missing dependent components are also described in the literature~\cite{fischer2020dependency} and presented in our taxonomy through particular instances of missing nodes and missing nodelet managers.
Physical unit misconfigurations within ROS have also been quantitatively and qualitatively studied through projects on GitHub~\cite{ore2017physical,canelas2024physunits}.
Issues related to URDF files presented in this study, through the scope of Xacro XML language, are also a source of misconfigurations in prior work~\cite{albergo2022xacro}.
ROSDiscover~\cite{timperley2022rosdiscover} and ROSInfer~\cite{duerschmid2024ROSInfer} identify misconfigurations related to the structural composition and behavioral interaction between components, covering certain misconfigurations within our Naming, Parameters and Messages, and Timing categories.
Finally, the ROBUST project ~\cite{timperley2019188,timperley2024robust}, a large-scale study of bugs in ROS, evidenced different types of misconfigurations related to missing runtime dependencies (e.g., nodes and configuration files), dangerous defaults (e.g., missing setting a parameter value), namespaces misconfigurations, and name mismatches.
Unlike prior work, we focus on how developers misconfigure their systems and which techniques are available to address them.

Configuration errors are not specific to robotic systems, and prior surveys found these in other configurable systems~\cite{xu2015survey}.
For instance, some open source storage systems are prone to inconsistency errors of parameter values and value inconsistencies~\cite{yin2011configerrors}, similar to the incorrect parameter and dependent parameter errors we encountered, respectively.
Similarly, Android manifests are also a source of incorrect attribute names and values~\cite{jha2017android}.
Finally, configuration errors in databases arise according to the data types used and the ranges of accepted values for the configuration parameters~\cite{xu2013misconfig}.

\section{Discussion}
\label{sec:discussion}

Misconfigurations are a critical concern in robotic systems, as these lead to unintended and potentially dangerous behavior.
Our study and literature review identified a gap in the ability of state-of-the-art analysis tools to cover the space of ROS misconfigurations.
In this section, we discuss some of the requirements that future analysis tools need to satisfy to improve the detection of the different categories of misconfigurations.


\paragraph{\textbf{Misconfiguration analysis must work with ROS's domain-specific languages and dialects.}} %

Building ROS Systems requires changing configuration elements distributed in multiple different configuration formats (e.g., Navigation Stack, MoveIt!, URDF).
Through our study, we observe that these formats are a source of misconfiguration (\keyword{Embedded DSL}).
As manually tracking many component configurations across different file formats and ensuring their consistency is challenging, automated techniques must detect misconfigurations throughout these files.
Through our literature review, we observe that only one technique explicitly treats these DSLs as first-class entities as part of its verification~\cite{kim2022robofuzz}. 

One avenue to address this concern is considering the DSL configurations in analysis tools.
Current analysis tools do not explicitly consider the semantics of the configurations within the different embedded DSLs.
A future direction in improving the detection of misconfigurations is to incorporate this knowledge into analysis tools to enhance their capabilities.
Alternatively, these separate configuration formats and files could be merged into a single analysis, verifying that configurations are correctly integrated across DSLs.

\paragraph{\textbf{Misconfiguration analysis require information about the robot's physical environment, hardware, and intended application to reliably detect misconfigurations}} %

As cyber-physical systems interact with the real world, correctly defining robot configurations depends on the context in which the system is used.
For instance, in \Cref{subsec:taxonomy-contextual}:\keyword{Contextual}, we identified misconfigurations arising from the lack of knowledge when changing configurations that depend on the environment, hardware, and type of application.
For instance, the positioning of sensors in the hardware, indoor and outdoor environments, the frequency, quality, and size of images provided by the sensors, and the type of robotic system are all factors found in this work that impact the configuration of software components.
If analysis tools intend to improve their verification, they must consider this contextual information.
However, the physical environment and hardware information is often missing, as our literature review found that only 6 of \npapersrelevants consider this information to optimize the configuration values.

Future work can provide application, physical environment, and system hardware information to analysis tools to improve their verification through two possible approaches: domain-specific languages and artifact mining.
Domain-specific languages have been successfully applied in other domains to verify system properties and improve code quality~\cite{kosar2016dsl,ketkar2024lightweight}.
Introducing a DSL, allows developers to specify properties regarding the context in which the system is executed (e.g., whether it is executed indoors or outdoors).
When existing, contextual information can be obtained by analyzing artifacts (e.g., Phys~\cite{kate2018phys}, ROSDiscover~\cite{timperley2022rosdiscover}, SA4U~\cite{taylor2022sa4u}), and inspecting the information these dialects provide.

\paragraph{\textbf{Static analysis is not sufficient to detect all misconfigurations.
Tools must be able to analyze run-time behavior.}} %

As executing the system is expensive, time-consuming, and possibly dangerous, it is ideal to detect these misconfigurations prior to system execution using static analysis tools.
While static analysis techniques perform great work in reducing the cost of detecting misconfigurations, these are still bounded by the limited context understanding, which is not provided in ROS-based systems, and have scalability issues, being challenging to analyze large codebases~\cite{landi1992static}.
Detecting some misconfigurations requires complex runtime behavior information not available to static analysis tools.
For instance, a set of parameter values may need to be corrected according to the system's execution (\keyword{Incorrect Parameters}),\footnote{\url{https://answers.ros.org/question/30235/}}
the incorrect calibration of the system is only detectable when executing the cameras (\keyword{Calibration}),\footnote{\url{https://answers.ros.org/question/10975/}} or compute issues may arise when hardware actively interacts with the real world (\keyword{Contextual}).\footnote{\url{https://answers.ros.org/question/195186/}}

Future analysis tools can improve their detection of misconfigurations by augmenting the static checking with properties.
Runtime behavior information could be used to generate test cases, monitor runtime configurations, or use machine learning techniques to predict potential misconfigurations based on contextual information and the system's execution.
These different approaches for analysis are currently used in techniques observed in our literature review and present a promising research direction.
For instance, HAROS~\cite{santos2018harospbt} generates monitors to track properties during runtime, SA4U~\cite{taylor2022sa4u} instruments the source code to obtain runtime information to help detect physical unit misconfigurations, and Swarmbug~\cite{jung2021swarmbug} performs multiple executions of the system while removing environment configuration variables to detect configurations responsible for the buggy behaviors.

Furthermore, future research can focus on system properties defined using specification languages.
Properties in these domain-specific languages can help monitor misconfigurations that are only detectable dynamically while interacting with the real world.

\section{Concluding Remarks}

In this work, we conduct an empirical study to categorize the misconfigurations that occur within ROS and determine the extent to which existing analysis tools cover those misconfigurations.
We find \nsubcategories categories of misconfiguration, of which \ncategoriesnotaddressed are found not to be addressed and \ncategoriesaddressed are partially addressed by existing tools.
Through this study, we identify promising areas for future research, outline requirements for future analysis tools, and, through our taxonomy, identify where detailed datasets are needed to develop tools to detect specific categories of misconfiguration.

\begin{acks}
    This work was supported by Fundação para a Ci\^encia e Tecnologia (FCT) in the LASIGE Research Unit under the ref. (UIDB/00408/2020, UIDP/00408/2020 and EXPL/CCI-COM/1306/2021), the CMU Portugal Dual PhD program (SFRH/BD/151469/2021), and NSF-USDA-NIFA \#2021-67021-33451. The authors would like to thank Bogdan Vasilescu and the Squareslab group for their feedback on this work.
\end{acks}

{
\bibliographystyle{ACM-Reference-Format}
\bibliography{references}
}
\end{document}